\documentclass[twocolumn,aps,showpacs,prc,superscriptaddress]{revtex4-1}
\usepackage[dvips]{epsfig}
\usepackage{graphicx}
\usepackage{amsmath}    

\begin{document}

\title{Fission fragment mass distributions in reactions populating $^{200}$Pb\\}

\author{A. Chaudhuri} 
\author {A. Sen}
\author {T. K. Ghosh} \email{E-mail: tilak@vecc.gov.in}
\author{K. Banerjee}
\author{Jhilam Sadhukhan}
\author{S. Bhattacharya} \thanks{Raja Ramanna Fellow}
\author {P. Roy}
\author {T. Roy}
\author{C. Bhattacharya}
\author{Md. A. Asgar}
\author{A. Dey}
\author {S. Kundu}
\author {S. Manna}
\author {J. K. Meena}
\author {G. Mukherjee}
\author {R. Pandey}
\author {T. K. Rana}
\author {V. Srivastava} 
\affiliation{Variable Energy Cyclotron Centre,  1/AF  Bidhan  Nagar,
  Kolkata  700064, India.}
	\author {R. Dubey}
	\author {Gurpreet Kaur}
	\affiliation{Department of Physics, Panjab University, Chandigarh 160014, India}
	\author {N. Saneesh}
	\author {P. Sugathan}
\affiliation {Inter University Accelerator Centre, Aruna Asaf Ali Marg, New Delhi 110067, India.}
\author{P. Bhattacharya}  
	\affiliation{Saha Institute of Nuclear Physics,  1/AF  Bidhan  Nagar, Kolkata  700064, India.}
\date{\today}

\begin{abstract}

The fission fragment mass distributions have been measured in the reactions  $^{16}$O + $^{184}$W and $^{19}$F + $^{181}$Ta populating the same compound nucleus $^{200}$Pb$^{*}$ at similar excitation energies. It is found that the widths of the mass distribution increases monotonically with excitation energy, indicating the absence of quasi-fission for both reactions. This is contrary to two recent claims of the presence of quasi-fission in the above mentioned reactions.

\end{abstract}


\maketitle

\section{INTRODUCTION}

Recently, the nucleus $^{200}$Pb has been studied widely, both theoretically and experimentally, to unravel the role of entrance channel dynamics on fusion process. Shidling \textsl{et al.} \cite{shidling1,shidling2} measured  the evaporation residue (ER) cross sections and gamma multiplicity distributions for $^{16}$O + $^{184}$W and $^{19}$F + $^{181}$Ta reactions leading to the same compound nucleus $^{200}$Pb$^{*}$. Both the systems under consideration have a charge product $Z_{P}$.$Z_{T}$ $<$ 700 (where Z$_{P}$ and Z$_{T}$ are projectile and target atomic numbers, respectively), much lesser than the approximate threshold value ($\geq$ 1600) for the onset of entrance channel dependence as per Swiatecki's dynamical model \cite{swiatecki1, swiatecki2}. Although the value of the entrance channel mass asymmetry ($\alpha = |A_T-A_P|/(A_T+A_P)$; $A_T, A_P$ being the target and projectile mass numbers) of the two systems (0.84 and 0.81, respectively) are similar, they are on either side of the Businaro-Gallone critical mass asymmetry ($\alpha_{BG}$ = 0.837) \cite{abe}. The measured (normalised) ER cross section and moments of gamma multiplicity distribution of the system $^{16}$O + $^{184}$W were found to be \cite{shidling1} significantly enhanced as compared to those of the other system  $^{19}$F + $^{181}$Ta at higher excitation energies, indicating entrance channel effects. As the reduction of ER yield in the reaction $^{19}$F + $^{181}$Ta, as compared to $^{16}$O + $^{184}$W  reaction, was correlated with a selective suppression of contributions of higher spin events in the former, the authors \cite{shidling1} attributed it to be due to the onset of  pre-equilibrium fission \cite{rama1,rama2} in the more symmetric system. 

Nasirov \textsl{et al.} \cite{nasirov} claimed that the observed reduction in ER cross-section ($\sigma_{ER}$), mentioned above \cite{shidling1}, at higher energies for the reaction $^{19}$F + $^{181}$Ta could be due to incorrect estimation of fusion cross-section ($\sigma_{fus}$). It was pointed out \cite{nasirov} that in the reconstruction  of $\sigma_{fus}$ \cite{shidling1} from fission like fragment yields, the contributions of quasi-fission and fast fission, which cause hindrance to complete fusion, were not properly identified and subtracted from the measured fission yield, leading to overestimation of $\sigma_{fus}$ and thereby a lowering of the normalised ($\sigma_{ER}/\sigma_{fus}$) ER yield.  Indeed, from the theoretical analysis performed in the framework of the dinuclear system and advanced statistical models \cite{dns}, Nasirov \textsl{et al.} showed that the magnitude of hindrance to complete fusion was different in the two systems - more for $^{19}$F + $^{181}$Ta as compared to $^{16}$O + $^{184}$W. It was interesting to note that
the calculation of Nasirov \textsl{et al.} predicted a dramatic increase in quasi-fission and fast-fission with increase in energy for the $^{19}$F + $^{181}$Ta reaction. Another recent fusion calculation \cite{Rajni} using dynamical cluster-decay model (DCM) and Wong model however suggest that, while $^{19}$F + $^{181}$Ta data can be explained well without incorporation of any quasi-fission, the presence of quasi-fission may not be ruled out in the case of $^{16}$O + $^{184}$W. The prevailing ambiguity has prompted us to have  a serious relook into the problem through a different experimental observable, fission fragment mass distribution, which has already been established to be a robust tool for direct detection of the presence/absence of quasi-fission in a nuclear reaction.

Variation of the width of the fragment mass distribution with excitation energies is known to be a sensitive probe for studying quasi-fission \cite{kaushik, myPLB, rgthomas}. As the statistical fission of the compound nucleus is expected to proceed through an unconditional mass symmetric fission barrier, the fission fragment mass distribution is symmetric around A$_{CN}$/2 (where A$_{CN}$ is the compound nucleus mass number), if fine structures due to shell effect are discounted for. Since shell effects are expected to be washed out at these excitation energies under consideration \cite{Abhirup}, the mass distributions should be symmetric with a smooth increase in width (or standard deviation $\sigma_{m}$) of the distribution with excitation energy \cite{AbhirupRapid}. Quasi-fission is a competing dynamical process which proceeds through a mass asymmetric conditional fission barrier, making the fragment mass distribution asymmetric. The mass distribution is also expected to be asymmetric for fast fission that occurs for the composite system when the angular-momentum-dependent fission barrier  becomes extremely small. However, the contribution of fast fission is negligibly small for our measured energy range \cite{nasirov}.Thus, an admixture of statistical fission events and quasi-fission will result in larger width of the mass distribution and the width of the mass distribution is expected to increase if there is an enhancement of quasi-fission with change in the excitation energy. Therefore, any sudden change in the width of the mass distribution would indicate departure from full equilibration, while onset of mass asymmetry or an increase in width of mass distribution would be a strong signal of quasi-fission.

In this work, we report the fission fragment mass distributions of the two aforementioned reactions, $^{19}$F + $^{181}$Ta and $^{16}$O + $^{184}$W populating the  compound nuclei $^{200}$Pb$^{*}$ at similar excitation energies to look for the presence/absence of quasi-fission. No significant deviation was found between the two entrance channels and the monotonic increase in the variation of the width (standard deviation) of the mass distributions clearly indicate the absence of quasi-fission in either of the reactions.   

\section{EXPERIMENTAL METHOD}

The experiment was performed at the 15UD Pelletron accelerator facility of the Inter University Accelerator Centre (IUAC), New Delhi with pulsed  beam of $^{19}$F and $^{16}$O on enriched isotopes of $^{181}$Ta of thickness 200 $\mu$g/$cm^2$ with carbon backing of 20 $\mu$g/$cm^2$ and self supporting $^{184}$W of thickness 100 $\mu$g/$cm^2$ respectively. The width of pulse beam was 1.2 ns with a repetition rate of 250 ns. Fission fragments were detected with two large area X-Y position sensitive multi-wire proportional counters (MWPCs) \cite{myNIM}. The MWPCs were mounted on two rotatable arms, at expected folding angles for complementary fission fragments. For the mass distribution measurements, the centre of the forward detector was kept at a polar angle ($\theta$) = 75$^{\circ}$  and the backward detector at $\theta$ = 74$^{\circ}$ on either side of the beam axis. The forward detector was placed at 41 cm from the centre of the target and the backward detector was placed at a lesser distance of 29 cm from the target so as to ensure complete coverage of complementary fission fragments.  The detectors were operated at a pressure of 3 torr of iso-butane gas. Operating the detector at low pressure improves the timing resolution and at the same time makes the detectors almost transparent to elastic and quasi-elastic particles.  For each event we measured the time of flight difference of the complementary fragments through the fast anode pulses, the X-Y coordinates of the fragments on the detector ($\theta,\phi$) and the energy loss of the fragment inside the detector. From these measurements, we extracted the masses of the correlated fission events and the momentum transferred to the fissioning system. Two silicon surface barrier detectors were placed at $\pm$10$^{\circ}$ for beam flux monitoring and normalization using the collected elastic events. The Faraday cup was also used as a means to normalize the observed fission events.
\begin{figure}
\includegraphics*[scale=0.25, angle=0]{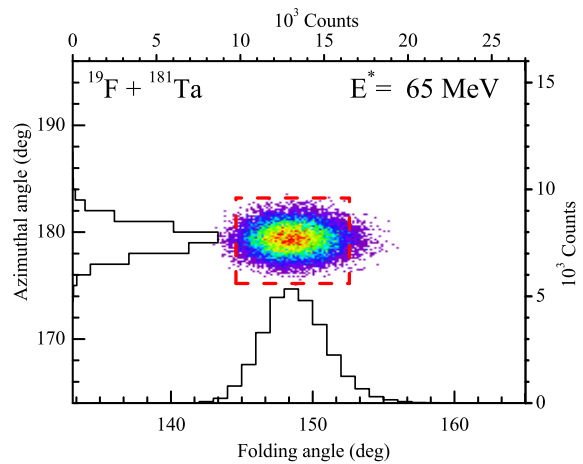}
\caption{\label{fig:fig1}~(Color online)Folding angle distributions of complimentary fission fragments for the system, $^{19}$F+ $^{181}$Ta at an excitation energy of 65 MeV. The rectangle indicates the gate used to select the FF events for mass determination.} 
\end{figure}

\section{DATA ANALYSIS}

A typical polar and azimuthal angle correlation plot of all fission fragments measured at an excitation energy of 65 MeV is shown in Fig. 1 for $^{19}$F + $^{181}$Ta reaction. The peak of the folding angle distribution is consistent with the value expected for complete transfer of momentum of the projectile. These distributions being symmetric in both $\theta$ and $\phi$ suggests that there is no admixture of transfer induced fission. However, the width of the polar and azimuthal angular correlations is enhanced due to both fission reaction kinematics and the spread due to post scission neutron emission from fragments. As inclusion of the latter is undesirable, events within a gate as shown in the figure were analyzed. We have checked that the analysis of the  data with narrower gates does not affect the width of the mass distributions within the error bar. We achieved mass resolution $\sim 4 u$ in the present experimental set up.

While transfer induced fission is minimal, the inherent properties of the detectors ensure that fragments are well separated from elastic and quasi-elastic reaction channels in the time correlation and energy loss spectra. The difference of the time of flights, polar and azimuthal angles, momentum, and the recoil velocities were used to determine the masses of fission fragments. The procedure has been described in details earlier papers \cite{myNIM,mythesis,myrapid}. 

\begin{figure}
\includegraphics*[scale=0.4, angle=0]{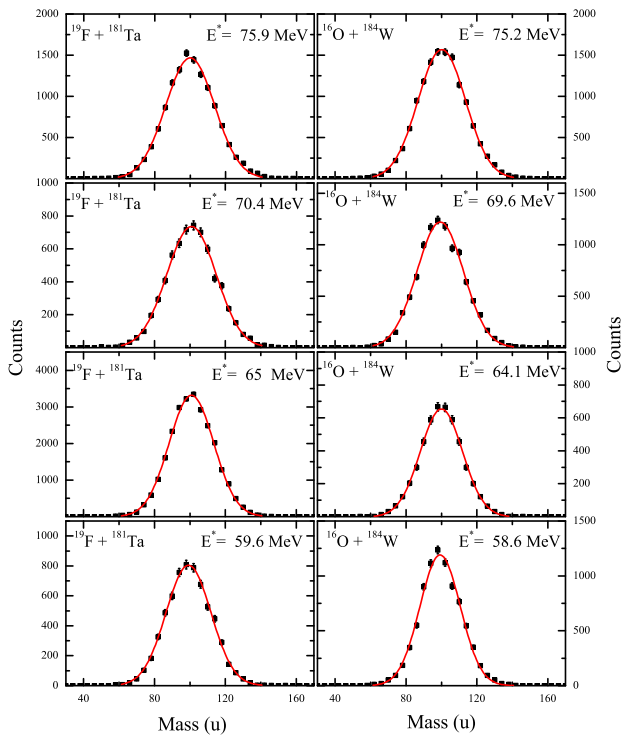}
\caption{\label{fig:fig2}~(Color online) Experimental mass distributions of fission fragments for the reactions $^{19}$F + $^{181}$Ta (left) and $^{16}$O + $^{184}$W (right) at different excitation energies. They were fitted by single Gaussian shown by full (red) lines.} 
\end{figure}

\section{RESULTS and DISCUSSIONS}

The typical mass distributions of fission fragments for both the reactions at similar excitation energies are shown in Fig. 2. The mass distributions at all energies are symmetric in nature and can be fitted with a single Gaussian, as shown by solid (red) line, with peak close to the half of the combined target-projectile mass. The presence of quasi-fission usually leads to asymmetry in the mass distribution (in the form of increased yields near target and projectile masses) and thereby causes additional broadening of the distribution.   It will be apparent from the following discussions that there was no significant admixture of an asymmetric distribution in the measured mass distributions.

The variation of standard deviations $\sigma_{m}$(u) of the fitted experimental mass distributions plotted as a function of  excitation energy is shown in Fig. 3. It shows that $\sigma_{m}$(u) increases smoothly with excitation energy for both the reactions and there is no anomalous large scale deviation in $\sigma_{m}$(u) at higher excitation energies between the two the reactions $^{19}$F + $^{181}$Ta and $^{16}$O + $^{184}$W. In the inset of Fig. 3, we show the ratio of cross sections of (quasi-fission + fast fission) and fusion for the above two reactions as predicted by Nasirov \textsl{et al.} \cite{nasirov}. It indicates significant dominance of quasi-fission and fast fission in the former as compared to the later. Qualitatively, this should have been reflected in the widths of their respective mass distributions (in the form of anomalous increase of the width in the former with respect to the later)- which was not observed at all.  Unfortunately,  no model exists at present that can predict the quantitative change in width of the mass distribution with changing quasi-fission fraction; however, it may be mentioned that the present tool is sensitive enough to detect an admixture of $\sim 5\%$ quasi-fission in a reaction \cite{kaushik}. 
     
In case of statistical fission of the equilibrated compound nucleus, the variance ($\sigma_m^2$)of the fission fragment mass distribution is a linear function of the nuclear temperature at saddle point. Dashed (blue) line in Fig. 3. shows the calculated standard deviation from statistical theory \cite{Back96}  following the relation $\sigma_m$ = $\sqrt{\frac{T}{k}}$, where $T$ is the temperature at the scission point, $k$ is the stiffness parameter for the mass asymmetry degree of freedom. Since the saddle and the scission point temperatures for the systems under investigations are very similar \cite{PRC07}, we used the saddle point temperature to calculate the standard deviation of the mass distributions. The temperature of the nucleus at the saddle point can be estimated as 
\begin{equation} 
T= \left[\frac{E^{\star}_{CN}- B_f(l) - E_{pre}- E_{rot}}{a}\right]^{1/2}
\end{equation}
where $E^{\star}_{CN}$ is the excitation energy of the compound nucleus, $B_f(l)$ is the angular momentum dependent fission barrier height, $E_{rot}$ is the rotational energy of the CN at the saddle point calculated according to the finite range rotating liquid-drop model \cite{sierk}, $E_{pre}$ is the energy carried out by pre-fission neutrons, which is estimated from the empirical systematic \cite{Itkis_temp} and $a$ is the nuclear level density parameter. 

\begin{figure}
\includegraphics*[scale=0.32, angle=0]{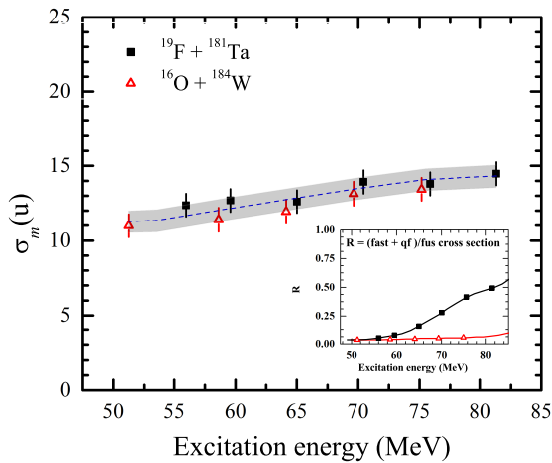}
\caption{\label{fig:fig3}~(Color online) Variation of the standard deviation $\sigma_{m}$(u) of the fitted symmetric mass distribution with excitation energy. The calculated widths are shown in (blue) dashed line, the shaded region indicate uncertainties in calculation (see text). The predicted \cite{nasirov} variation of the sum of the fast fission and quasi-fission cross sections (normalized with respect to the fusion cross sections) with excitation energy are presented in the inset. } 
\end{figure}

It is to be mentioned here that the variance of the mass distribution also has a weak dependence on the mean square average value of angular momentum $<l^{2}>$ \cite{Itkis_temp}:
\begin{equation}
\sigma_{m}(u)= \sqrt{(T/k + \beta<l^{2}>)}
\end{equation}
with the value of the constant $\beta \sim 0.05$ \cite{PRC07}. The angular momentum ($<l^{2}>$) of the CN  was calculated by CCFULL code \cite{hagino}. A value of the inverse stiffness parameter $1/k = (98.1 \pm 15.1) u^2/MeV$ fitted the data well  and was found to be consistent with the comprehensive compilation of the data \cite{PRC07,Itkis_k}. The uncertainty in calculation of $\sigma_{m}$ due to the uncertainty in the compiled value of inverse stiffness parameter $1/k$ is shown by the shaded region in Fig 3. It is evident that mass variances of both the systems followed the same trend within the limits of uncertainty. The admixture of fast and quasi-fission as predicted in the theoretical calculation \cite{nasirov} for the reaction $^{19}$F + $^{181}$Ta (as shown in the inset of Fig 3), which gradually increases to as high as $\sim$50\% with the increase in excitation energy,  should have made the trend of variation of the standard deviation ($\sigma_{m}$) of mass distribution drastically different as compared to that for the $^{16}$O + $^{184}$W reaction at higher excitation energy. On the contrary, the present measurement clearly indicates that, both the reactions follow fusion fission path and there was no appreciable difference in the fusion dynamics for the  two reactions in the measured excitation energy range.

\section{THEORETICAL CALCULATION OF FISSION MASS WIDTH}

The present non observation of any appreciable quasi-fission in either of the two reactions is further confirmed by our theoretical calculation. We solved two-dimensional Langevin equations with elongation ($c$) and mass-asymmetry ($\alpha$) as collective coordinates \cite{jhilam} to estimate the mass distributions for both the target-projectile combinations. The input angular momentum for each Langevin event is sampled from the corresponding CCFULL spin distribution. The Langevin equations are written as \cite{wada}
  \begin{align}
  \label{langevin}
  \frac{d p_i}{dt} &=
  -\frac{p_j p_k}{2} \frac{\partial}{\partial x_i}(\mathcal{M}^{-1})_{jk}
  - \frac{\partial F}{\partial x_i}\\ \nonumber
  & - \eta_{ij}(\mathcal{M}^{-1})_{jk}p_k
  + g_{ij}\Gamma_j(t),\\ \nonumber
  \frac{dx_i}{dt} &= (\mathcal{M}^{-1})_{ij}p_j,
  \end{align}
  where $x_i$ represents either of $c$ and $\alpha$ and $p_i$ is the associated conjugate momentum. The driving force $F$ is given by $F=U-(a-a_0)T^2$, where $U$ is the potential energy calculated from the finite-range liquid drop model \cite{sierk} and using rigid rotor values for moment of inertia, $a$ is the Ignatyuk's shape dependent level density parameter \cite{ignatyuk} with its value $a_0$ at the ground state deformation. The temperature $T$ is calculated from the available excitation energy $E^*$ using the relation $E^*=a_0T^2$. The inertia tensor $\mathcal{M}$ is evaluated from the Werner-Wheeler prescription \cite{davies}. The expression for the dissipation tensor $\eta$ with proper reduction factor as prescribed in references \cite{jhilam,karpov}, was used in our calculation. In equation \ref{langevin}, $g_{ij}\Gamma_j(t)$ is  the random (Langevin) force with $\Gamma_j(t)$ being a time-dependent stochastic variable with a Gaussian distribution, and $g_{ij}$  is the random-force strength tensor. The time-correlation property of the random force is assumed to follow the relation $\langle\Gamma_k(t)\Gamma_l(t')\rangle = 2\delta_{kl}\delta(t-t')$. The strength of the random force is related to the dissipation coefficients through the fluctuation-dissipation theorem: $\sum_kg_{ik}g_{jk}=\eta_{ij}k_{\rm B}T$.
\begin{figure}
\includegraphics*[scale=0.37, angle=0]{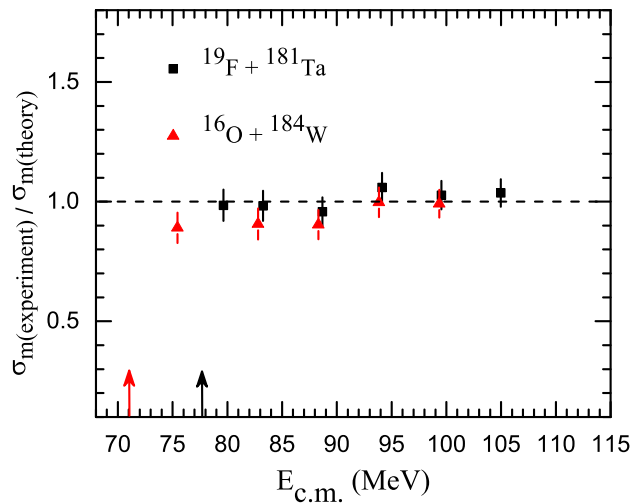}
\caption{\label{fig:fig4}~(Color online) Variation of the ratio of the experimental (measured) and theoretically calculated standard deviation $\sigma_{m}$(u) of the fitted symmetric mass distribution with c.m. energy. The Coulomb barriers for the $^{19}$F + $^{181}$Ta (77.8 MeV) and $^{16}$O + $^{184}$W (71 MeV) reactions are shown by arrows. The dashed line (constant value = 1) is a guide to the eye.} 
\end{figure}

We calculated $\sigma_m$ from the theoretically obtained mass distributions. The ratios of experimental and theoretical $\sigma_m$ values are plotted in Fig. 4. There is a good overall agreement between the experimental results and the theoretically obtained values as the points in Fig. 4 are very close to $1.0$ clearly indicating the absence of quasi-fission in $^{19}$F + $^{181}$Ta reaction.

\section{MASS ANGLE CORRELATION}

In the case of quasi-fission reactions, a correlation between the fragment mass and angle exist as the composite system breaks before completing a full rotation \cite{Nishio}. In order to look for the possible signature of quasi-fission, in Fig. 5, the mass angle correlation for the reaction $^{19}$F + $^{181}$Ta is plotted at a representative excitation energy of 75.9 MeV at which a large amount of quasi-fission cross-section were predicted \cite{nasirov}. However, no significant correlation of fragment mass with angle was observed indicating the absence of quasi-fission.

\begin{figure}
\includegraphics*[scale=0.45, angle=90]{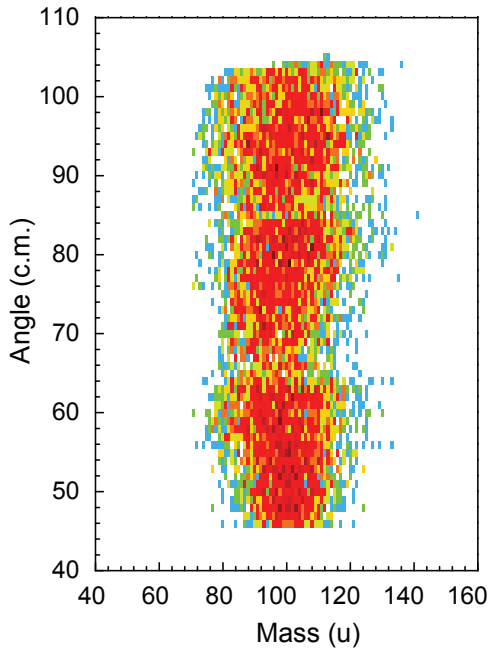}
\caption{\label{fig:fig5}~(Color online) Measured mass angle distributions of the fission fragments in the reaction $^{19}$F + $^{181}$Ta at excitation energy =75.9 MeV } 
\end{figure}

\section{SUMMARY AND CONCLUSION}

It is thus apparent from the present study that, for the two systems discussed above, the fusion dynamics pathways are almost identical and there is no indication of any substantial contribution from non-equilibrium reaction mechanisms like quasi-fission. Incidentally, in earlier studies of admixture of quasi-fission \cite{kaushik}, fission fragment mass distribution was found to be a sensitive tool even in cases where other probes like fragment angular distribution or pre-scission neutron multiplicity were not conclusive. The absence of any deviation from statistical model predicted width \cite{nasirov} of the mass distribution even at the highest excitation energy, where quasi-fission contributions should be more significant if present, leads us to infer that quasi-fission is not significantly present in either of the two reactions.

This, however, leaves the following question remain unanswered; if not quasi-fission, then what is the cause of the lowered ER yields reported in reference \cite{shidling1}? If pre-equilibrium fission is, as suggested \cite{shidling1}, the mechanism behind the above phenomenon, it is not clear whether the present probe will be sensitive to it (in other words, whether pre-equilibrium fission is also associated with wider mass distribution or not). In absence of any available theoretical prediction in this regard, we refrain from making any definite comment at this juncture and lay stress on the need for advanced theoretical models to distinguish between the subtle features of various non-compound fission processes such as fast-fission, quasi-fission and pre-equilibrium fission.  

In conclusion, the present study of fission fragment mass distribution did not find any signature of quasi-fission for the reactions $^{19}$F + $^{181}$Ta and $^{16}$O + $^{184}$W. The finding is contrary to two recent claims of presence of quasi-fission in the above mentioned reactions. More calculations are required to understand the fusion dynamics of pre actinide nuclei.

\section{ACKNOWLEDGEMENTS}
We are thankful to the staff members of the IUAC Pelletron for providing good quality of pulsed beam required for the experiment. Thanks to HYRA group and IUAC target laboratory for providing the $^{181}$Ta target. Thanks are due to Dr. Avazbek Nasirov and Dr. Santanu Pal for illuminating discussions. A.C., A.A., T.R. and V.S. acknowledge the financial support received from the Department of Atomic Energy, Government of India. A.D acknowledges with thanks the financial support provided by the SERB, Department of Science and Technology, Government of India; and S. B. acknowledges with thanks the financial support received from the Department of Atomic Energy, Government of India.

\end{document}